\documentclass[useAMS,usenatbib,usegraphicx]{mn2e}
\usepackage{natbib}
\usepackage{color}
\usepackage{mathrsfs}
\usepackage{times,xspace,amssymb}
\usepackage{url}
\usepackage{hyperref}
\usepackage{amsmath}
\usepackage{ulem}
\usepackage{float}
\usepackage{multirow}

\bibpunct{(}{)}{;}{a}{}{,}

\def\mbh{M_{\bullet}}
\def\msun{M_{\odot}}
\def\ergs{\rm erg\,s^{-1}}
\def\kms{\rm km\,s^{-1}}
\def\cmc{\rm cm^{-3}}
\def\funit{\rm erg\,cm^{-2}\,s^{-1}}
\def\rs{R_{\rm s}}
\def\ms{M_{\rm s}}
\def\mp{m_{\rm p}}
\def\me{m_{\rm e}}
\def\sigt{\sigma_{\rm T}}
\def\rhog{\rho_{\rm g}}
\def\ng{n_{\rm g}}
\def\vs{v_{\rm s}}
\def\pt{P_{\rm t}}
\def\et{E_{\rm t}}

\def\tcool{t_{\rm cool}}
\def\et{E_{\rm t}}

\def\lin{L_{\rm in}}
\def\lff{L_{\rm ff}}
\def\tff{t_{\rm ff}}

\def\lic{L_{\rm IC}}
\def\lsyn{L_{\rm syn}}

\def\tdyn{t_{\rm dyn}}
\def\tic{t_{\rm IC}}

\def\lp{L_{\rm p}}
\def\tp{t_{\rm p}}

\def\ledd{L_{\rm Edd}}
\def\tedd{\tau_{\rm Edd}}

\def\lagn{L_{\rm AGN}}
\def\lkin{L_{\rm kin}}
\def\lnt{L_{\rm nt}}

\def\ent{\epsilon_{\rm nt}}
\def\mh{M_{\rm halo}}
\def\mbg{M_{\rm bulge}}
\def\md{M_{\rm DM}}
\def\mgal{M_{\rm gal}}
\def\rvir{R_{\rm vir}}
\def\rdisk{R_{\rm disk}}
\def\cn{c_{_N}}
\def\tvir{T_{\rm vir}}

\def\vin{v_{\rm in}}
\def\th{t_{\rm H}}
\def\re{r_{\rm e}}

\def\fg{f_{\rm g}}
\def\fd{f_{\rm d}}

\def\mtot{M_{\rm tot}}
\def\cd{C_{\rm d}}
\def\ch{C_{\rm h}}
\def\fin{f_{\rm in}}
\def\fagn{f_{_{\rm AGN}}}
\def\sigt{\sigma_{\rm T}}

\def\xib{\xi_{_{\rm B}}}
\def\ent{\epsilon_{\rm nt}}

\def\uagn{U_{\rm AGN}}
\def\ucmb{U_{\rm CMB}}
\def\uph{U_{\rm ph}}

\def\gmax{\gamma_{\rm max}}
\def\gmin{\gamma_{\rm min}}
\def\gb{\gamma_{\rm b}}
\def\kacc{\xi_{\rm acc}}
\def\rl{R_{\rm L}}
\def\vsyn{\nu_{\rm syn}}
\def\tsyn{t_{\rm syn}}
\def\vic{\nu_{_{\rm IC}}}

\def\mc{\multicolumn}
\def\mr{\multirow}

\def\mnras{MNRAS}
\def\apj{ApJ}
\def\aj{AJ}

\def\aap{A\&A}
\def\apjl{ApJL}
\def\apjs{ApJS}

\def\physrep{Phys. Rep.}
\def\araa{Ann. Rev. A \& A}
\def\nat{Nature}

\title[Probing gaseous galaxy halos]
{Probing the gaseous halo of galaxies through non-thermal emission from AGN-driven outflows}
\author[X. Wang and A. Loeb]
{Xiawei Wang \thanks{E-mail: xiawei.wang@cfa.harvard.edu}
and Abraham Loeb
\\
Department of Astronomy, Harvard University, 60 Garden St., Cambridge, MA 02138, USA}
\begin{document}
\date{Accepted 2015 July 17. Received 2015 July 15; in original form 2015 June 16}

\pagerange{\pageref{firstpage}--\pageref{lastpage}} \pubyear{2015}

\maketitle

\label{firstpage}

\begin{abstract}
Feedback from outflows driven by active galactic nuclei (AGN) can affect the distribution and properties of the gaseous halos of galaxies.
We study the hydrodynamics and non-thermal emission from the forward outflow shock produced by an AGN-driven outflow. 
We consider a few possible profiles for the halo gas density,
self-consistently constrained by the halo mass, redshift and the disk baryonic concentration of the galaxy.
We show that the outflow velocity levels off at $\sim 10^3\,\kms$ within the scale of the galaxy disk. 
Typically, the outflow can reach the virial radius around the time when the AGN shuts off.
We show that the outflows are energy-driven, consistently with observations and recent theoretical findings.
The outflow shock lights up the halos of massive galaxies across a broad wavelength range.
For Milky Way (MW) mass halos, 
radio observations by \textit{The Jansky Very Large Array} (\textit{JVLA}) and \textit{The Square Kilometer Array} (\textit{SKA}) and infrared/optical observations by \textit{The James Webb Space Telescope} (\textit{JWST}) and \textit{Hubble Space Telescope} (\textit{HST}) can detect the emission signal of angular size $\sim 8''$ from galaxies out to redshift $z\sim5$.
Millimeter observations by \textit{The Atacama Large Millimeter/submillimeter Array} (\textit{ALMA}) are sensitive to non-thermal emission of angular size $\sim 18''$ from galaxies at redshift $z\lesssim1$,
while X-ray observations by \textit{Chandra}, \textit{XMM-Newton} and \textit{The Advanced Telescope for High Energy Astrophysics} (\textit{ATHENA}) is limited to local galaxies ($z\lesssim 0.1$) with an emission angular size of $\sim2'$.
Overall, the extended non-thermal emission provides a new way of probing the gaseous halos of galaxies at high redshifts.
\end{abstract}
\begin{keywords}
shock waves -- galaxies: active -- galaxies: haloes -- quasars: general -- radio continuum: general. 
\end{keywords}
%
%
\section{Introduction}
Outflows from active galactic nuclei (AGN) regulate black hole (BH) growth \citep{1998A&A...331L...1S,2005Natur.433..604D}
and may quench star formation \citep{2005MNRAS.361..776S,2008ApJS..175..356H} in galaxies.
A great amount of observational evidence has demonstrated the presence of AGN-driven outflows, including observations of absorptions in quasars \citep{2007ApJ...665..990G, 2009ApJ...696.1693F, 2009ApJ...706..525M, 2011MNRAS.418.2032V, 2013MNRAS.436.3286A, 2014MNRAS.442..784Z, 2015A&A...577A..37A},
multiphase outflows in nearby ultraluminous infrared galaxies (ULIRGs) \citep{2011ApJ...729L..27R, 2011ApJ...733L..16S, 2014A&A...562A..21C, 2015arXiv150301481F, 2015Natur.519..436T} and quasars \citep{2015arXiv150603096C, 2015arXiv150600614G},
and post-starburst galaxies \citep{2011Sci...334..952T}.
The velocity of AGN-driven outflows can reach $\sim 10^3\,\kms$ on galaxy scale, indicating that the outflows are likely to propagate into the halos of galaxies while the AGN is active.
Here we propose to use AGN-driven outflows as a probe of the halo gas in galaxies.

Halo gas has been identified in multiphases (see review by Putman et al. 2012): cold neutral hydrogen detected as high velocity clouds \citep{2005A&A...440..775K, 2005A&A...432..937W, 2007AJ....134.1019O, 2012ApJ...758...44S},
warm gas ($T\sim 10^{4-5}$ K) discovered in deep $\rm H\alpha$ emission line surveys
\citep{2003ApJ...597..948P, 2012MNRAS.424.2896L},
warm-hot gas ($T\sim 10^{5-6}$ K) detected in absorption \citep{2009ApJS..182..378W, 2009ApJ...690.1558P, 2011ApJ...731...14S, 2013MNRAS.433.1634M, 2014MNRAS.441..886F, 2014MNRAS.444.1260F}
and hot gas ($T\sim 10^6$ K) inferred from X-ray observations in emission and absorption \citep{2013ApJ...772...97B, 2013ApJ...770..118M, 2015ApJ...804...72B}.
The presence of warm-hot and hot halo gas, extending out to the virial radius, is of particular interest since the hot gas is postulated to host a significant fraction of baryons in the galaxy \citep{2006MNRAS.370.1612K}.
However, the detailed properties and the origin of the extended and diffuse hot halo gas remain uncertain since there is little evidence for its existence around spiral galaxies \citep{2012ARA&A..50..491P}.
The detection of halo gas out to virial radius scale is difficult and the extent to which the outflows impact the properties of the halo gas remains uncertain.
Therefore, it is important to study the interaction between AGN-driven outflows and surrounding gas on different scales as a probe of the properties of the diffuse hot halo gas and the effectiveness of the feedback mechanism.

In galaxies with a weaker AGN where the energetics of AGN activity and star formation are comparable, it remains unclear whether outflows are dominated by AGN or supernovae (SN) \citep{2015arXiv150405209H}.
In this paper, we focus on AGN-driven outflows.
First, our model assumes spherical symmetry, which is more justified for AGN-driven outflows since they are launched at the center of the galaxy whereas SN-driven outflows are distributed throughout the entire disk.
More importantly, as shown later in the paper, the strongest emission signal comes from more massive galaxies where AGN feedback is thought to dominate.

Previous work on the dynamics of and non-thermal emission from galactic outflows has made simple assumptions about the total gravitational mass and the gaseous environment in which the outflow propagates \citep{2001ApJ...556..619F, 2003ApJ...596L..27K, 2011MNRAS.415L...6K, 2012MNRAS.425..605F, 2015MNRAS.447.3612N},
and limited the evolution of the outflows to galactic disk scales \citep{2010ApJ...711..125J, 2012MNRAS.425..605F, 2015MNRAS.447.3612N, 2015arXiv150405209H}.
In this paper, we explore different gas density profiles in galaxy halos and examine the non-thermal emission from the forward shock plowing into the ambient medium in details. 
We predict the multiwavelength spectrum and detectability of the non-thermal emission and discuss how the outflow shock and halo gas affect each other.
We propose a new way to probe the gaseous halo using the non-thermal emission from the outflow shocks as they travel through the ambient medium in the galaxy and halo.

The paper is organized as follows. In \S~\ref{sec:model}, we describe our model for the halo and gas distribution.
In \S~\ref{sec:hydro}, we analyze the hydrodynamics of AGN-driven outflows.
In \S~\ref{sec:emission}, we calculate the non-thermal emissions from shocks produced by outflows.
In \S~\ref{sec:result}, we show numerical results for representative cases and discuss physical significance.
Finally in \S~\ref{sec:summary}, we summarize our results and discuss their implications.
%
%
\section{Model description}
\label{sec:model}
We approximate the galaxy and halo as spherically symmetric. The environment into which the outflow propagates is decribed below.
Here we discuss properties of spherical outflows driven by fast nuclear wind \citep{2010ApJ...711..125J, 2015arXiv150305206K}.
The predicted radio emission from outflow shocks as discussed in \S~\ref{sec:emission} is fainter than the radio synchrotron emission from relativistic jets in a small subset of all active galaxies \citep{2014ARA&A..52..589H}.
\subsection{Mass profile of host galaxy}
We assume that the density distribution of the galaxy in which the outflow is initially embedded
follows the NFW profile \citep{1996ApJ...462..563N}:
\begin{equation}
\rho_{_{\rm NFW}}(R)=\rho_0(1+z)^3\frac{\Omega_m}{\Omega_m(z)}
\frac{\delta_c}{\cn x(1+\cn x)^2} \;,
\end{equation}
where $\rho_0=3H_0^2/8\pi G$ is the critical density today, $H_0$ is the Hubble constant today, $G$ is the gravitational constant,
$x=R/\rvir$, $\cn$ is the concentration parameter which is roughly given by:
$\cn\approx25(1+z)^{-1}$, $\Omega_m=0.3$. 
$\delta_{\rm c}$ is given by 
$\delta_c=\Delta_c c_N^3/[3(\ln(1+\cn)-\cn/(1+\cn))]$, where $\Delta_c\approx18\pi^2$.
$\Omega_{\rm m}(z)$ can be expressed as
$\Omega_m(z)=\Omega_m(1+z)^3/[\Omega_m(1+z)^3+\Omega_{\Lambda}]$,
where $\Omega_{\Lambda}=0.7$.
$\rvir$ is the virial radius, written as $\rvir=0.78\,h^{-2/3}\left[\Omega_{\rm m}\Delta_{\rm c}/18\pi^2\Omega_{\rm m}(z)\right]^{-1/3}\,M_{\rm h,8}^{1/3}/(1+z/10)\,\rm kpc$,
where $h=(H_0/100\,\kms)$ is the Hubble parameter and $\mh=10^{8}M_{\rm h,8}\msun$ is the halo mass.
We obtain the total mass of the galaxy and dark matter halo within a radius of $R$ by
$\int\,4\pi R^2\rho_{_{\rm NFW}}(R)\,dR$, which gives:
\begin{equation}
\begin{split}
\md(R)=&\rho_0(1+z)^3\frac{\Omega_m}{\Omega_m(z)}\frac{\delta_c\rvir^3}{\cn^3}\\
&\left[\ln(1+\cn x)-\frac{\cn x}{1+\cn x}\right]\; .
\end{split}
\end{equation}
We estimate the BH mass $\mbh$ self-consistently by the following steps \citep{2015ApJ...806..124G}.
First we obtain the total stellar mass in the galaxy $M_{\star}$ determined by $\mh$ \citep{2010ApJ...710..903M}:
\begin{equation}
M_{\star}=M_{\star,0}\frac{\left(\mh/M_1\right)^{\gamma_1}}
{\left[1+\left(\mh/M_1\right)^{\beta}\right]^{(\gamma_1-\gamma_2)/\beta}}\;,
\end{equation}
where $\log (M_{\star,0}/\msun)=10.864$, $\log (M_1/\msun)=10.456$, $\gamma_1=7.17$,
$\gamma_2=0.201$ and $\beta=0.557$.
There is no specific bulge mass $M_{\rm bulge}$ for a given halo mass \citep{2013ARA&A..51..511K}.
Numerical simulation \citep{2014MNRAS.441..599B} suggests that the bulge-to-total stellar mass ratio $\rm B/T=$$M_{\rm bulge}/M_{\star}$ is roughly uniformly distributed from $0$ to $1$.
This ratio for the MW is $\sim 0.15$ \citep{2014AAS...22333604L}.
Additionally, Fisher \& Drory (2011) suggest that $\sim 25\%$ of all local stellar mass is in bulges and elliptical galaxies.
We then adopt a particular value of $\rm B/T$ ratio to be $\sim 30\%$ in our calculation and multiply this value by $M_{\star}$ to get $M_{\rm bulge}$ to illustrate some examples.
There is likely to be only ellipticals in high mass halos, so it is justified to take a fixed B/T ratio for these systems.
We also verify that modifying B/T ratio only results in a difference within a factor of $4$.
This variation can be cancelled out by the uncertainty in the fraction of AGN's luminosity injected into the medium as discussed later in the paper.
Finally, we obtain the BH mass $\mbh$ by \citep{2013ApJ...764..184M}:
\begin{equation}
\log(\mbh/\msun)=8.46+1.05\log\left[\frac{\mbg}{10^{11}\msun}\right]\;.
\end{equation}
The underestimation of the $\mbh$ correlations could be an issue for the most massive BHs \citep{2013ARA&A..51..511K} but should not affect our results on the emission from the outflow shocks in the more common galaxies with $\mbh\ll10^9\,\msun$.
\subsection{Gas density profile}
We assume that the gas takes up a fraction $\fg$ of the total mass of the dark matter in a galaxy.
We adopt a cosmic mean baryon fraction, which is $\fg\sim 16\%$ \citep{2013ApJS..208...19H}.
A fraction of the baryons $\fd$ is concentrated in the disk of the galaxy,
and the disk radius $\rdisk$ is taken to be $\sim 4\%$ of the virial radius $\rvir$ \citep{2015arXiv150307481S}.

\begin{figure}
\includegraphics[angle=0,width=\columnwidth]{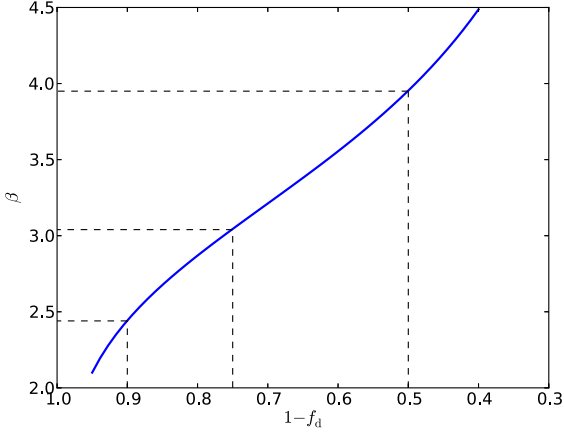}
\caption{\label{pl_index}Power-law index $\beta$ of the halo gas density profile as a function of the baryon fraction of the halo ($1-\fd$).
The dashed lines correspond to values of $\fd=0.1$, $0.25$ and $0.5$, which we have
taken into numerical calculation in the following sections.
}
\end{figure}

Our first prescription for the gas density distribution is a broken power-law profile, given by:
\begin{eqnarray}
\rho_{\rm pl}(R)=
\begin{cases}
\cd\,R^{-\alpha}		& (R\le\rdisk) \\
\ch\,R^{-\beta}		& (\rdisk<R\le\rvir) \\
\end{cases}
\end{eqnarray}
where $\alpha$ and $\beta$ are the power-law indices in the disk and halo component, respectively.
We assume an isothermal sphere for the gas within the disk component and fix $\alpha=2.0$ in our calculation.
The constants in the density profile $\cd$ and $\ch$ can be constrained by
the baryon mass budget in the disk component and in total.
Consequently, $\beta$ is soley dependent on $\fd$.
The constraint on $\beta$ by $\fd$ is shown in Fig.\ref{pl_index},
where we find that when $\fd\sim 0.25$, $\beta\sim3$, indicating that the gas in the halo approximately follows the NFW profile.
From the broken power-law density profile, we estimate the gas number density at $50-100$ kpc to be $10^{-5}-10^{-4}\,\cmc$, 
which is consistent with numerical simulations \citep{2015arXiv150404620S} and observations \citep{2015ApJ...804...72B} of the hot halo gas distribution.

The second profile we consider for the halo gas density distribution is analogous to that of galaxy clusters, written as \citep{2015ApJ...798L..20P}:
\begin{equation}
\rho_{\rm clu}(R)=\Gamma \fg A\frac{(R/s)^{2\Gamma-2}}
{R/r_{\rm s}\left[1+(s/r_{\rm s})(R/s)^{\Gamma}\right]^2}\; ,
\end{equation}
where $A=\rho_0\delta_{\rm c}$ is the scale parameter, $s=\rvir$, $r_{\rm s}=s/c_N$ is the scale radius and $\Gamma$ is the jump ratio.
The density profile recovers to a scaled NFW profile for $\Gamma=1$.
%
%
\section{Hydrodynamics}
\label{sec:hydro}
We assume spherical symmetry for the outflow and the ambient medium. 
Fast wind with velocity $\sim 0.1c$ is injected into the medium, as inferred from observations of broad absorption lines in quasars \citep{2013MNRAS.436.3286A}.
The wind drives an outer forward shock into the ambiet medium accelerating the swept-up material and an inner reverse shock into the wind decelerating itself, separated by a contact discontinuity \citep{2015arXiv150305206K}.

The equation of motion of the shell is given by \citep{2001ApJ...556..619F, 2012MNRAS.425..605F}:
\begin{equation}
\frac{d^2 \rs}{dt^2}=\frac{4\pi\rs^2}{\ms}(\pt-P_0)-\frac{G\mtot}{\rs^2}-\frac{\vs}{\ms}\frac{d\ms}{dt}\, ,
\end{equation}
where $G$ is the gravitational constant, and $\rs$, $\vs$ and $\ms$ are the radius, velocity and mass of the swept-up shell, respectively. 
$\mtot$ is the total gravitational mass inside $\rs$ that impedes the expansion of the wind bubble, written as $\mtot=\md+\mgal+\mbh+\ms/2$, composed of the mass of dark matter $\md$, galaxy $\mgal$, the central BH $\mbh$, and the self-gravity of the shell. 
The shell mass, $\ms$, satisfies,
\begin{equation}
\frac{d\ms}{dt}=4\pi\rhog\rs^2\vs\, ,
\end{equation}
where $\rhog$ is the ambient gas density profile in the galaxy.

Hydrostatic equilibrium gives the temperature in the ambient medium $T_0$:
\begin{equation}
\frac{dT_0}{dR}=\frac{G\mtot\mp}{kR^2}-\frac{T_0}{\ng}\frac{d\ng}{dR}\,,
\end{equation}
where $\mp$ is the proton mass, $k$ is the Boltzman constant and $\ng$ is the number density profile of the ambient gas. 
At virial radius $\rvir$, $T_0$ reaches virial temperature $\tvir=\mu\mp v_{\rm c}^2/2k$ where $\mu=0.5$ is the mean molecular weight of fully ionized gas and
 $v_{\rm c}$ is the circular velocity, given by $v_{\rm c}=\left(G\mh/\rvir\right)^{1/2}$.
The ambient thermal pressure is given by $P_0=\ng kT_0$.

The thermal pressure in the shocked wind $\pt$ declines due to radiative energy losses and work done on the ambient gas by the expansion, at a rate:
\begin{equation}
\frac{d\pt}{dt}=\frac{\Lambda}{2\pi\rs^3}-5\pt\frac{\vs}{\rs}\, ,
\end{equation}
where $\Lambda$ is the heating and cooling function, composed of energy injection from the central source and different physical cooling processes in the shocked wind region:
\begin{equation}
\label{eq:11}
\Lambda=\lin-\lff-\lic-\lsyn-\lp\, .
\end{equation}
Energy is continuously injected into the shocked wind during the quasar's lifetime,
taken to be the e-folding time $\tedd\approx4.5\times10^7\rm yrs$ \citep{2001ApJ...547...12M},
with a rate of $\lin$, which is assumed to be a fraction of the AGN's bolometric luminosity $\fin\lagn$.
Observations infer $\fin$ to be $\sim1\%-5\%$ \citep{2013MNRAS.436.3286A, 2014A&A...562A..21C} and we adopt $\fin=5\%$ in our calculation.
We assume that $\lagn$ is a fraction $\fagn$ of the Eddington luminosity $\ledd=1.38\times10^{38}(\mbh/\msun)\;\ergs$,
and adopt $\fagn=0.5$ in our calculation \citep{2009ApJ...697.1656S}.

The last four terms in the right hand side of Eqn.~\ref{eq:11} account for radiative cooling.
$\lff$ is the radiative cooling rate via free-free emission in the shocked wind.
$\lic$ decribes cooling via inverse Compton (IC) scattering off photons in the quasar's radiation field and the cosmic microwave background (CMB).
$\lsyn$ represents synchrotron cooling rate.
$\lp$ refers to the cooling of protons through Coulomb collisions with the electrons.
The cooling rate can be expressed as $\mu\et/t_{\rm c}$, where $\et=2\pi\rs^3\pt$ is the thermal energy in the shocked wind and $t_{\rm c}$ is the timescale corresponding to different cooling processes.
The total emissivity of free-free emission is given by \citep{1979rpa..book.....R}: 
$\epsilon_{\rm ff}=1.4\times10^{-22}\,T_{\rm e,10}^{1/2}\,n_{\rm e,0}^{2}\,\bar{g}_{_{\rm B}}$,
where $\bar{g}_{_{\rm B}}$ is the Gaunt factor, $T_{\rm e,10}=(T_{\rm e}/10^{10}\rm\, K)$ and $n_{\rm e,0}=(n_{\rm e}/1\,\cmc)$ are the electron temperature and number density, respectively.
The corresponding cooling timescale is $\tff=\frac{3}{2}kT_{\rm e}/\epsilon_{\rm ff}=
4.69\times10^8\, T_{\rm e,10}^{1/2}\,n_{\rm e,0}^{-1}\,\bar{g}_{_{\rm B}}^{-1}\,\rm yr$.
The IC cooling time of relativistic electrons of energy $E_{\rm e}$ in soft photon radiation field can be written as \citep{2015arXiv150305206K}:
$\tic=3\me^2c^3/8\pi \sigt \uph E_{\rm e}$,
where $\sigt$ is the Thomson scattering scross section and $\uph$ is the energy density of soft photons, including AGN photons with energy density $\uagn=\lagn/4\pi\rs^2 c$ and CMB photons with energy density $\ucmb\approx4.2\times10^{-13}(1+z)^4\,\rm erg\,cm^{-3}$.
Here we consider the most efficient IC cooling limit and thus leave out non-relativistic electrons, of which the IC cooling time can be significantly longer \citep{2012MNRAS.425..605F}.
We obtain the temperature in the shocked wind by the Rankine-Hugoniot jump condition
$T_{\rm e}\approx 3\mu\mp\vin^2/16 k$.
The synchrotron cooling timescale is given by $\tsyn=1.6\times10^{12}\,B_{-6}^{-2}\,T_{e,10}^{-1}\,\rm yr$, where $B_{-6}=(B/10^{-6}\,\rm G)$.
If two-temperature plasma effect is taken into account \citep{2012MNRAS.425..605F}, then the proton cooling timescale $\tp$ can be expressed as: $\tp\approx1.4\times10^9\,R_{\rm s,kpc}^2\,L_{\rm AGN,46}^{-1}\,v_{\rm s,3}^{2/5}\,v_{\rm in,0.1}^{8/5}\,\rm yr$,
where $v_{\rm s,3}=(\vs/10^3\,\kms)$ and $L_{\rm AGN,45}=(L_{\rm AGN}/10^{45}\,\ergs)$.
%
\section{Non-thermal emission}
\label{sec:emission}
Next we discuss the non-thermal emission from the outflow shock as it propagates in the ambient medium \citep{2015MNRAS.447.3612N}.
\subsection{Synchrotron emission}
As the forward shock plows through the ambient medium supersonically, 
a broken power-law distribution of non-thermal electrons $N(\gamma)\,d\gamma\propto\gamma^{-p}\left(1+\gamma/\gb\right)^{-1}
$ is generated via Fermi acceleration in the shock to produce non-thermal emission, where $p$ is the power-law index. 
$\gb$ is the break Lorentz factor, which is obtained by equating the dynamical timescale $\sim \rs/\vs$ and the cooling timescale $3\me c/4(U_{\rm B}+\uagn+\ucmb)\sigt\gamma$.
This gives $\gb=3\me c\vs/4\sigt\rs(U_{\rm B}+\uagn+\ucmb)$, where $\me$ is the electron mass, $\sigt$ is the Thomson scattering cross section and $U_{\rm B}=B^2/8\pi$ is the energy density of the magnetic field.
We assume that the total non-thermal luminosity is a fraction of the kinetic energy
of the swept-up material, written as
$\lnt=\ent\lkin\approx\frac{1}{2}\ent\dot{\ms}\vs^2$.
We calibrate the magnetic field energy density as a fraction $\xib$ of the thermal energy behind the shock in what follows supernova (SN) remnants \citep{1998ApJ...499..810C}, giving:
\begin{equation}
U_{\rm B}=\xib nkT\; .
\end{equation}
Observations of radio emitting bubbles from a radio-quiet quasar imply $p\sim 2$ \citep{2015ApJ...800...45H}.
By fitting the radio flux from bubbles at $\sim 10$ kpc, we obtain $\ent\sim 5\%$.
Coefficients $\xib$ can be estimated from observations of late-time radio emission from relativistic jets associated with tidal disruption events \citep{2013ApJ...763...84B}, synchtrotron emission from shocks between jet and circumnuclear medium \citep{2012MNRAS.420.3528M} as well as from an analogy with SN remnants \citep{1998ApJ...499..810C}.
These observations imply $\xib\sim 0.1$.

\begin{figure}
\includegraphics[angle=0,width=\columnwidth]{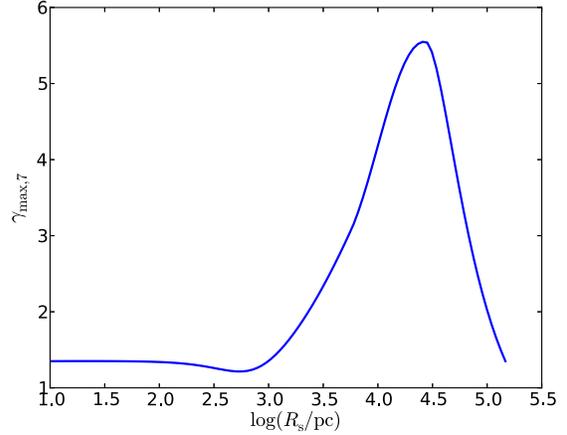}
\caption{\label{gmax}The maximum Lorentz factor of non-thermal electrons $\gmax$ in unit of $10^7$ as a function of outflow shock radius.
We fix $\mh=10^{12}\,\msun$, $\fd=0.25$ and $z=1.0$ as a representative example.
}
\end{figure}

Finally, we calculate the synchrotron emission following the standard formula from \citep{1970ranp.book.....P, 1979rpa..book.....R}.
The emission and absorption coefficients are given by:
\begin{equation}
j_{\nu}^{\rm syn}=c_{1}B\int^{\gmax}_{\gmin} F(x)
N(\gamma)\,d\gamma\; ,
\end{equation}
\begin{equation}
\alpha_{\nu}^{\rm syn}=-c_{2}B\frac{1}{\nu^{2}}\int^{\gmax}_{\gmin} \gamma^{2}
\frac{d}{d\gamma}\left[\frac{N(\gamma)}{\gamma^{2}}\right]
F(x)\,d\gamma\; ,
\end{equation}
where 
$c_1=\sqrt{2}e^{3}/4\pi \me c^{2}$, $c_2=\sqrt{2}e^{3}/8\pi \me^{2}c^{2}$,
$F(x)\equiv x\int^{\infty}_{x} K_{5/3}(\xi)\,d\xi$ and
$K_{5/3}(x)$ is the modified Bessel function of $5/3$ order.
The maximum Lorentz factor $\gmax$ is given by the tighter constraint of equaling the acceleration timescale $\kacc\rl c/\vs^2$ \citep{1987PhR...154....1B} to either dynamical or cooling timescale,
where $\kacc\sim 1$ and $\rl=\gamma\me c^2/eB$ is the Larmor radius.
We plot $\gmax$ in unit of $10^7$ as a function of outflow shock radius $\rs$ for $\mh=10^{12}\,\msun$, $\fd=0.25$ and $z=1.0$ as a representative example, shown in Fig.\ref{gmax}.
$\gmax$ varies within a factor of $\sim 5$ as a result of simultaneously decreasing $\vs$ and soft photon energy density with increasing $\rs$.
We take the minimum Lorentz factor $\gmin\sim1$ in our calculation.
The synchrotron emission peaks at a frequency of $\vsyn=4.2\times10^{14}\,B_{-6}\,\gamma_{7}^2\,\rm Hz$, where $\gamma_{7}=(\gamma/10^7)$.
\subsection{Inverse Compton scattering}
The soft photons includes those from the accretion disk and CMB.
The energy density of the AGN radiation field is $\uagn\approx2.8\times10^{-10}L_{\rm AGN,45}\,R_{\rm s,kpc}^{-2}\,\rm erg\,cm^{-3}$.
The CMB photons have an energy density of $U_{_{\rm CMB}}\propto(1+z)^4$,
which manifests themselves as a dominant source of IC scattering at high-redshift \citep{2004MNRAS.353..523C}.
The spectral energy distribution of quasars can be constrained by observations \citep{1994ApJS...95....1E, 2004MNRAS.351..169M, 2014MNRAS.438.2253S}.
For simplicity, we approximate it as a black body spectrum \citep{2015arXiv150408166I}.
We model the CMB photons as a black body with a spectrum peak frequency of $\nu_{_{\rm CMB}}\approx1.6\times10^{11}(1+z)$ Hz.
The peak of IC scattering of CMB photons takes place at a frequency of $\vic\approx \gamma^2\nu_{_{\rm CMB}}=1.6\times10^{25}\,\gamma_7^2(1+z)\,\rm Hz$.
The differential rate to produce high-energy photons with energy $\epsilon\me c^2$ is given by \citep{1968PhRv..167.1159J, 1990MNRAS.245..453C}:
\begin{equation}
Q(\epsilon)=\int\,d\epsilon_0\,n(\epsilon_0)\int\,d\gamma N(\gamma)\,K(\epsilon,\gamma,\epsilon_0)\;,
\end{equation}
where $\epsilon_0\me c^2$ is the soft photon energy, $\gamma\me c^2$ is the electron energy and $n(\epsilon_0)$ is the number density of soft photons.
$K(\epsilon,\gamma,\epsilon_0)$ is the Compton kernel, expressed as:
\begin{equation}
\begin{split}
K(\epsilon,\gamma,\epsilon_0)=\frac{2\pi\re^2 c}{\gamma^2\epsilon_0}
[&2\kappa\ln\kappa+(1+2\kappa)(1-\kappa)\\
&+\frac{(4\epsilon_0\gamma\kappa)^2}{2(1+4\epsilon_0\gamma\kappa)}(1-\kappa)]\;,
\end{split}
\end{equation}
where $\kappa=\epsilon/[4\epsilon_0\gamma(\gamma-\epsilon)]$.
The emission coefficient of IC scattering can be obtained by:
\begin{equation}
j_{\nu}^{\rm IC}=\frac{h}{4\pi}\epsilon Q(\epsilon)\;,
\end{equation}
where $h$ is the Planck constant.
%
\section{Numerical results}
\label{sec:result}
\begin{figure*}
\includegraphics[angle=0,width=2.0\columnwidth]{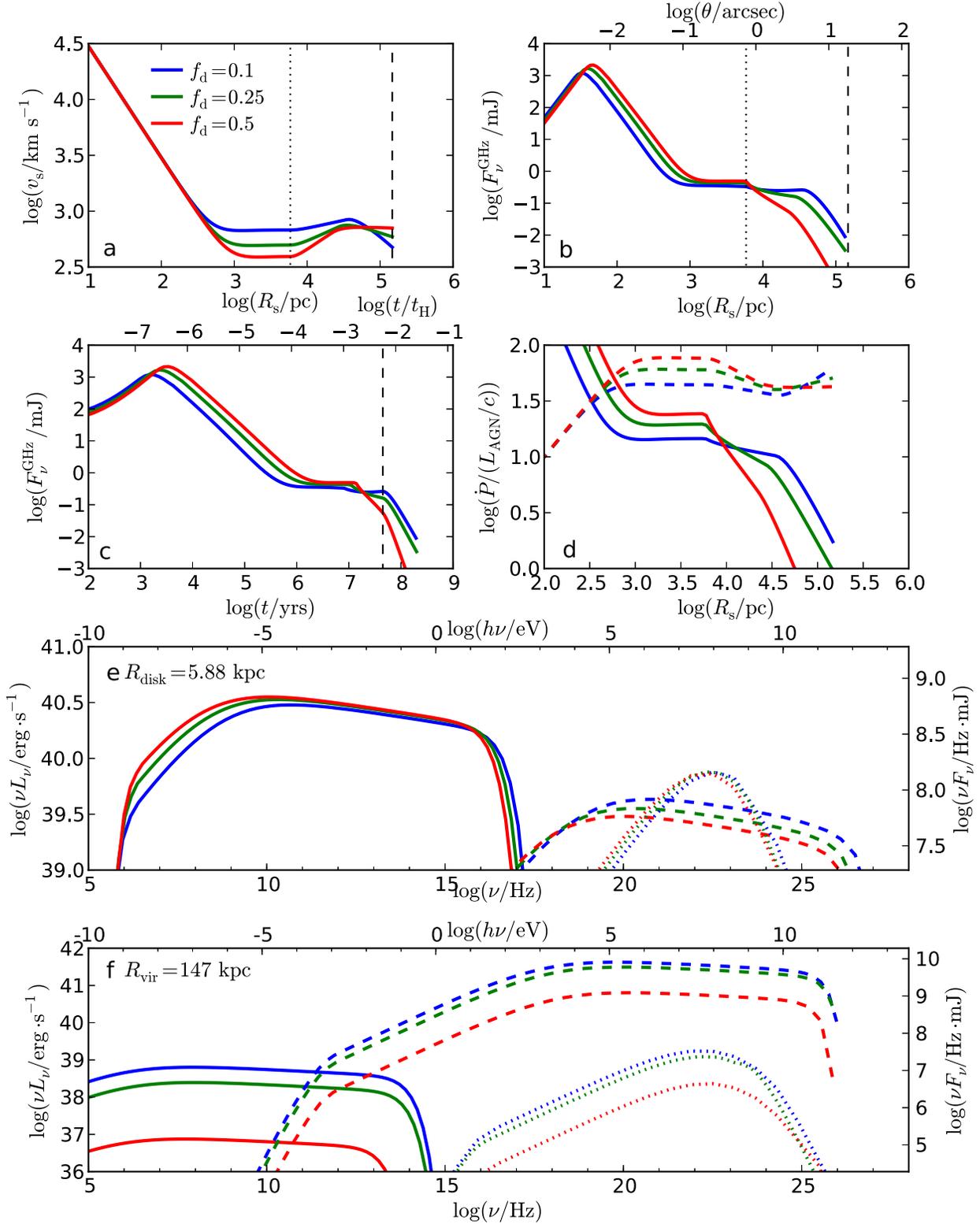}
\vglue -0.5cm
\caption{\label{fd}
Dependence of outflow hydrodynamics and emission on baryon fraction in the disk $\fd$. We fix $\mh=10^{12}\msun$ and $z=1.0$. 
Panel \textit{a} and \textit{b} show the shell velocity and radio synchrotron flux at 1 GHz as a function of radius.
The dotted and dashed vertical lines mark the position of $\rdisk$ and $\rvir$, respectively.
The upper x-axis of panel \textit{b} marks the angular diameter of the outflow shock.
Panel \textit{c} shows the radio synchrotron flux as a function of time. The dashed vertical line corresponds to the point when the AGN shuts off. 
Time is scaled to the Hubble time $\th$ on the upper x-axis.
Panel \textit{d} demonstrates the momentum flux boost of the shell. The solid lines represent the numerical result while the dashed lines correspond to predictions in the energy-driven regime.
Panel \textit{e} and \textit{f} illustrate snapshots of non-thermal emission power and flux at $\rdisk$ and $\rvir$, respectively. The solid, dashed and dotted lines correspond to synchrotron emission, IC scattering of accretion disk photons and CMB photons, respectively.
}
\end{figure*}

In Figures \ref{fd}--\ref{gc}, we show the dependence of outflow hydrodynamics solutions and emissions on $\fd$, $\mh$, $z$ and density profile formulation.
Since the gas distribution in the intergalactic medium (IGM) is uncertain, we restrict our calculation to halo scale within $\rvir$.

As shown in panel \textit{a} in Figures \ref{fd}--\ref{gc}, we find that the swept-up shell decelerates quickly to a roughly constant velocity of $\sim 10^{3}\,\kms$ in the disk. As it propagates outside the galaxy into the halo, the shell accelerates somewhat as a result of the tenuously distributed halo gas.
The evolution of the shell velocity is consistent with a self-similar solution,
where the shell radius is assumed to follow $\rs\propto t^{\delta}$ and $\vs\propto t^{\delta-1}$.
We express the gas power-law density profile generally as $\rho\propto R^{\gamma}$.
For $\gamma<3$, we obtain $\ms\propto R^{3-\gamma}$.
In the energy-conserving limit, we assume that $\sim 50\%$ of the injected energy goes to the kinetic energy of the swept-up material, $\lin t=\ms\vs^2$, and so we have $\delta=3/(5-\gamma)$.
For power-law index $\alpha=2$ in our model, $\delta=1$ and thus $\vs$ approaches a constant in the disk.
We can also verify that for halo component power-law index $\beta$, the outflow accelerates as $\beta>2$.
The acceleration stops as the quasar shuts off and
the thermal energy in the shocked wind $\et$ drives the expansion of the shell afterwards.
At this point, the outflow reaches the edge of the dark matter halo and is likely to continue to propagate into the IGM.

Panels \textit{b} and \textit{c} in Figures \ref{fd}--\ref{gc} show the radio flux as a function of shock radius and time, respectively.
We scale the time to the Hubble time $\th$, which is given by
$\th\equiv1/H(z)=H_0^{-1}\left[\Omega_{\rm m}(1+z)^3+\Omega_{\Lambda}\right]^{-1/2}$.
The chance of finding a galaxy with a given flux is $t/\th$.
We find that for $z\sim1$, about a few percent of the galaxy halos embed outflows reaching $\rvir$.
We also calibrate the angular diameter of the outflow shock, given by $\rs/D_{\rm A}$, where $D_{\rm A}$ is the angular distance.

We show snapshots of non-thermal emission taken at two milestones in panels \textit{e} and \textit{f}.
At the edge of the galaxy disk, the energy injection from the central source has an age of $\sim 10^7$ yrs.
At the virial radius, snapshots are taken at the dead quasar remnants with outflow approaching the edge of the dark matter halo on a timescale of $\sim 10^8$ yrs, which indicates that this population should be $\sim 10$ times more abundant.
At this point, the outflow no longer overlaps with the galaxy and there is no galaxy-bubble interactions.
We find that the outflows can reach the edge of the halo around the end of quasar's lifetime.
This feature indicates that AGN-driven outflows are most abundant during their passage through their host galaxy halo.

We summarize the detectability of this extended non-thermal emission in Table ~\ref{table:detectability}.
\begin{table*}
\begin{center}
{\footnotesize
\begin{tabular}{lllllllll}\hline\hline
            &\mc{2}{c}{$z=0.1$} & &\mc{2}{c}{$z=1.0$} & & \mc{2}{c}{$z=5.0$}\\ \cline{2-3}\cline{5-6}\cline{8-9}
\mr{2}{*}{Telescopes} & $F_{\nu}(R_{\rm disk})$ ; $F_{\nu}(R_{\rm vir})$   & \mr{2}{*}{detectability}    & & $F_{\nu}(R_{\rm disk})$ ; $F_{\nu}(R_{\rm vir})$  & \mr{2}{*}{detectability}  & & $F_{\nu}(R_{\rm disk})$ ; $F_{\nu}(R_{\rm vir})$  & \mr{2}{*}{detectability}
\\ 
   &\mc{1}{c}{(mJy)}  &  &  &\mc{1}{c}{(mJy)}   &    &  &\mc{1}{c}{(mJy)}  &
\\
\hline
JVLA		& 300 ; 0.8   &Yes ; Yes &   &1.0 ; $5\times10^{-3}$   &Yes ; Yes&   &$4\times10^{-3}$  ; $6\times10^{-4}$   &Yes ; Marginal\\
SKA        & 300 ; 0.8    &Yes ; Yes &   &1.0 ; $5\times10^{-3}$   &Yes ; Yes&   &$4\times10^{-3}$ ; $6\times10^{-4}$  &Yes ; Marginal\\
ALMA     & 0.5 ; $4\times10^{-3}$   &Yes ; Marginal &   &$5\times10^{-3}$ ; $7\times10^{-5}$   &Marginal ; No&   &$4\times10^{-5}$ ; $4\times10^{-4}$   &No ; No \\
JWST       & $3\times10^{-4}$ ; $5\times10^{-6}\,^{^{*}} $   &Yes ; Marginal &   &$5\times10^{-5}$ ;  $2\times10^{-6}\,^{^{*}}$   &Yes ; No&   &$4\times10^{-8}$ ; $2\times10^{-5}$   &No ; Yes  \\
HST       & $2\times20^{-4}$ ; $5\times10^{-6}\,^{^{*}}$    &Yes ; No &   &$3\times10^{-5}$ ; $2\times10^{-6}\,^{^{*}}$   &Marginal ; No&   &$4\times10^{-8}$ ; $2\times10^{-5}$   &No ; Marginal  \\
\hline
&$\nu F_{\nu}(R_{\rm disk})$ ; $\nu F_{\nu}(R_{\rm vir})$ &\mr{2}{*}{detectability}  &  &$\nu F_{\nu}(R_{\rm disk})$ ; $\nu F_{\nu}(R_{\rm vir})$   &\mr{2}{*}{detectability}  &   &$\nu F_{\nu}(R_{\rm disk})$ ; $\nu F_{\nu}(R_{\rm vir})$   &\mr{2}{*}{detecatbility}\\
  &\mc{1}{c}{($\funit$)}  &  &  &\mc{1}{c}{($\funit$)}  &  &  &\mc{1}{c}{($\funit$)}  &
\\
\cline{2-3}\cline{5-6}\cline{8-9}
XMM-Newton & $10^{-16}$ ; $2\times10^{-16}$    &Marginal ; Marginal &   &$5\times10^{-19}$ ; $4\times10^{-17}$   &No ; No&   &$7\times10^{-20}$ ; $10^{-17}$   &No ; No\\
ATHENA  & $10^{-16}$ ; $2\times10^{-16}$  &Yes ; Yes &   &$5\times10^{-19}$ ; $4\times10^{-17}$   &No ; Marginal&   &$7\times10^{-20}$ ; $10^{-17}$   &No ; No\\
Chandra    & $2\times10^{-17}$ ; $7\times10^{-16}$   &No ; Marginal &   &$5\times10^{-19}$ ; $7\times10^{-17}$   &No ; No&   &$6\times10^{-20}$ ; $10^{-17}$   &No ; No \\
NuSTAR  & $2\times10^{-17}$ ; $7\times10^{-16}$    &No ; No &   &$5\times10^{-19}$ ; $7\times10^{-17}$   &No ; No&   &$6\times10^{-20}$ ; $10^{-17}$   &No ; No\\

\hline
\end{tabular}
\caption{\label{table:detectability}Detectability of non-thermal emission from AGN-driven outflow shock.}
\vskip 0.1cm
\parbox{2.0\columnwidth}
{
Note: 
We choose $\mh=10^{12}\msun$ and $\fd=0.25$ as a representative example for a galaxy halo.
For radio, mm/sub-mm, infrared and optical observations, we provide values of $F_{\nu}(R_{\rm disk})$ and $F_{\nu}(R_{\rm vir})$, which correspond to non-thermal flux at the edge of the disk and halo respectively, in unit of mJy. For X-ray observation, we present $\nu F_{\nu}(R_{\rm disk})$ and $\nu F_{\nu}(R_{\rm vir})$, in unit of $\funit$.\\
The telescope detection limits are as follows: 
\begin{itemize}
\item[--]
\textit{The Jansky Very Large Array (JVLA)}: $\sim 1\mu$Jy for 1 $\sigma$ detection and 12h integration time at most bands \citep{nrao2014}.
\item[--]
\textit{The Square Kilometer Array (SKA-MID)}: $\sim 0.7\mu$Jy RMS sensitivity for 10h integration time \citep{2014arXiv1412.6942P}.
\item[--]
\textit{The Atacama Large Millimeter/submillimeter Array (ALMA)}:  At observating frequency $345$ GHz, the sensitivity $\sim 8.7\,\mu$Jy for 10h integration time is calculated by the  \textit{ALMA} Sensitivity Calculator (ASC) (https://almascience.eso.org/proposing/sensitivity-calculator).
\item[--]
\textit{The James Webb Space Telescope (JWST)}: sensitivity $\sim 10$ nJy for wavelength $1-3\,\mu$m and $\sim 30$nJy for wavelength $4-5\,\mu$m for 10$\sigma$ detection and $10^4$ s integration time \citep{stsci2013}.
\item[--]
\textit{Hubble Space Telescope (HST)}: sensitivity $\sim 40-50$ nJy for wavelength $0.6-1.5\,\mu$m for 10$\sigma$ detection and $10^4$ s integration time \citep{stsci2013}.
\item[--]
\textit{Chandra}: sensitivity of high resolution camera (HRC) $\sim 9\times 10^{-16} \funit$ covering energy range $0.08-10$ keV for 3$\sigma$ detection and $3\times 10^5$ s integration time \citep{chandra2014}.
\item[--]
\textit{XMM-Newton}: $\sim 3.1\times10^{-16}\,\funit$ in $0.5-2.0$ keV band \citep{2001A&A...365L..45H}.
\item[--]
\textit{Advanced Telescope for High Energy Astrophysics (ATHENA)}: $\sim 4\times 10^{-17}\,\funit$ in $0.5-2$ keV band in a $10^6$s deep field \citep{2012arXiv1207.2745B}.
\item[--]
\textit{Nuclear Spectroscopic Telescope Array (NuStar)}: $\sim2\times10^{-15}\,\funit$ in $6-10$ keV band for 3$\sigma$ detection and $10^6$ s integration time \citep{2013ApJ...770..103H}.
\end{itemize}
* The emission may be contaminated by scattered quasar light (see \S~\ref{sec:summary}).}}
\end{center}
\end{table*}
\subsection{Dependence on parameters}
\subsubsection{Disk mass fraction}
For a halo of mass $\mh=10^{12}\,\msun$ at $z=1.0$,
we choose three representative values of $\fd$ as motivated by observations \citep{2015MNRAS.448.1767C}.
We find that the shell velocity is not sensitive to $\fd$. The outflow reaches the edge of the halo 
around the time the energy injection discontinues. 
With a velocity of $\sim 500-10^3\,\kms$, the outflow is likely to propagate into the IGM.
The non-thermal radio flux at 1 GHz remains at $\sim 0.1$ mJy within the disk, independent of $\fd$. 
As the shell propagates into the halo, the non-thermal emission diminishes quicker in halos with higher $\fd$ as a result of more tenuous halo gas.
For $\fd=0.5$, the radio emission is $\sim 100$ times fainter than the other two cases and drops below the detection limit of \textit{JVLA} and \textit{SKA} before the outflow reaches $\rvir$.
Observationally, we can distinguish galaxies with high disk baryonic concentrations by
the faint emission from their outflows propagating in the halos.

\subsubsection{Halo mass}
We examine $\mh$ of $10^{11}\msun$, $10^{12}\msun$ and $10^{13}\msun$,
covering the full range from mid to high mass halos.
In lower mass halos,
the energy input into embedded outflows is much lower due to the 
self-consistent scaling relation between $\mbh$ and $\mh$.
The outflow shock decelerates quicker and may not propagate farther outside the galactic disks.
The short lifetime of outflows in low mass galaxy halos makes them less abundant.
Therefore, it would be observationally challenging to identify outflows from low mass halos
in terms of both emission intensity and recurrence rate.
At $z\sim1$, the emission is only detectable in radio band on galaxy scale with a flux $\sim 10\,\mu$Jy.
High mass galaxies produce AGN photons of higher energy density, making the detection more promising.
\subsubsection{Redshift}
The hydrodynamics of outflows is insensitive to $z$. 
Consequently, outflows reach the edge of its host galaxy and halo at similar velocities for different redshifts.
At low redshift $z\sim 0.1$, the non-thermal emission is detectable in multiwavelength from radio to X-ray.
For high-redshift galaxies at $z=5$, the non-thermal emission is dominated by IC scattering off CMB photons. 
The emission remains observable in the radio, infrared and optical bands on halo scale.
\subsubsection{Gas density profile}
We compare the broken power-law profile to the gas density profile of galaxy clusters.
We find that the outflow velocity and emission indistinguishable for these gas density profiles.
However, outflows can not reach the edge of the halo for galaxy clusters, excluding them from halo scale observations in these systems.

\begin{figure*}
\begin{centering}
\includegraphics[angle=0,width=2\columnwidth]{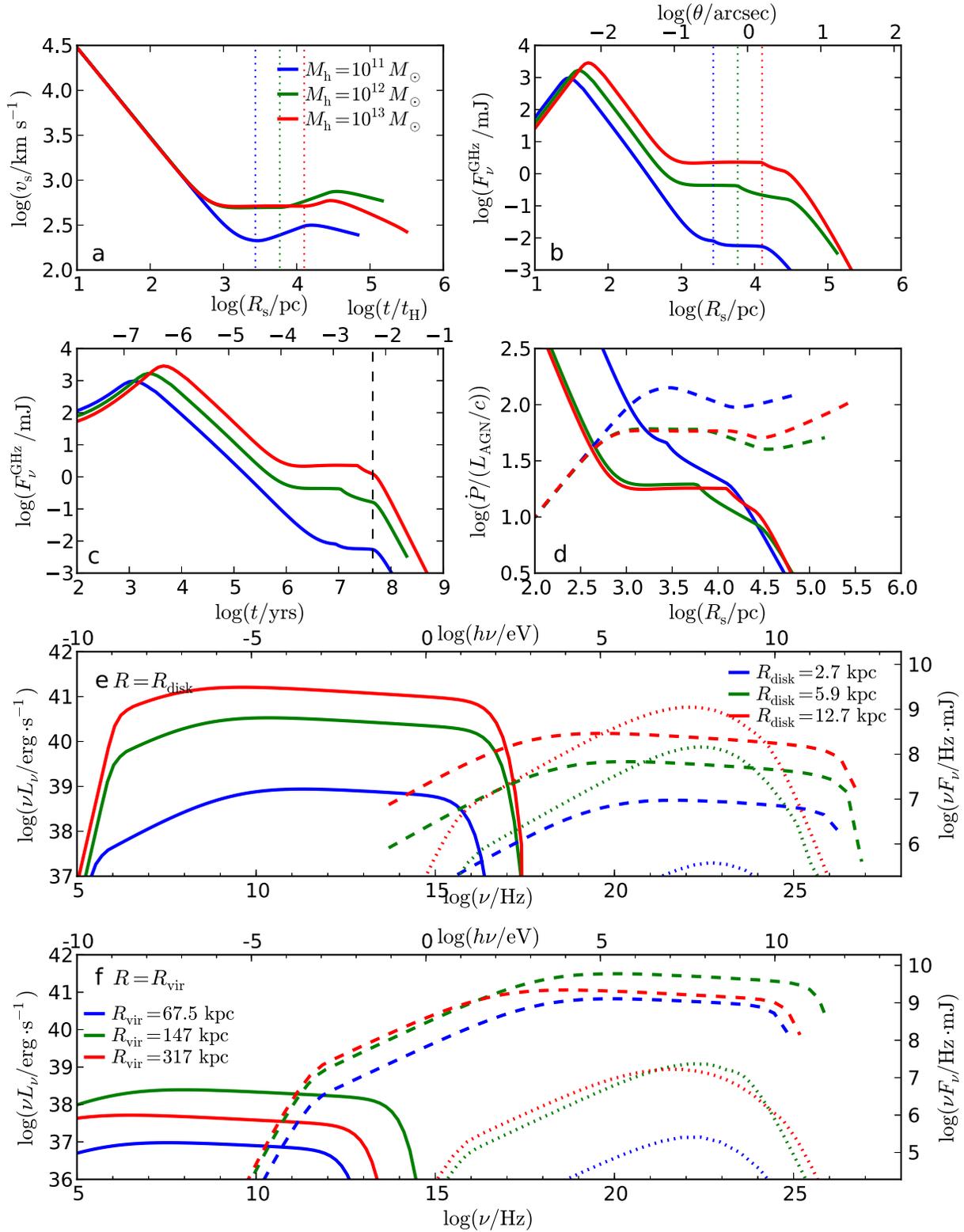}
\vglue -0.5cm
\caption{Dependence of outflow hydrodynamics and emission on halo mass $\mh$.
We fix $\fd=0.25$ and $z=1.0$. 
The configuration and physical significance of the subplots are the same as Fig.~\ref{fd}.
The dotted vertical lines marks the position of $\rdisk$ for the three cases in panel \textit{a} and \textit{b}.
}
\end{centering}
\end{figure*}
\begin{figure*}
\includegraphics[angle=0,width=2\columnwidth]{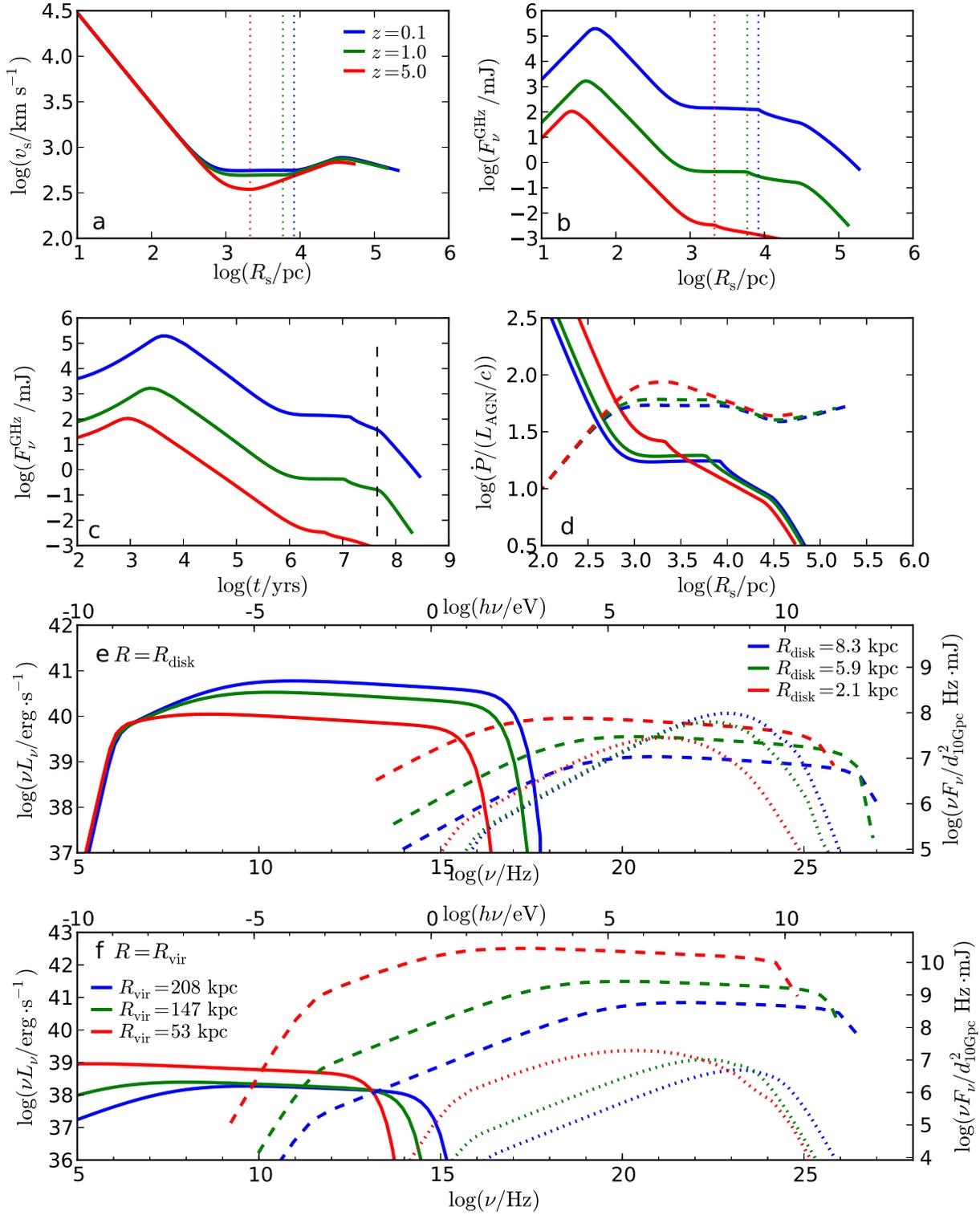}
\vglue -0.5cm
\caption{Dependence of outflow hydrodynamics and emission on redshift $z$.
We fix $\mh=10^{12}\msun$ and $\fd=0.25$. 
The configuration and physical significance of the subplots are the same as Fig.~\ref{fd}.
The dotted vertical lines marks the position of $\rdisk$ for the three cases in panel \textit{a} and \textit{b}.
The right-hand y-axis of panel \textit{e} and \textit{f} is scaled to a distance of $10$ Gpc.
}
\end{figure*}
\begin{figure*}
\includegraphics[angle=0,width=2\columnwidth]{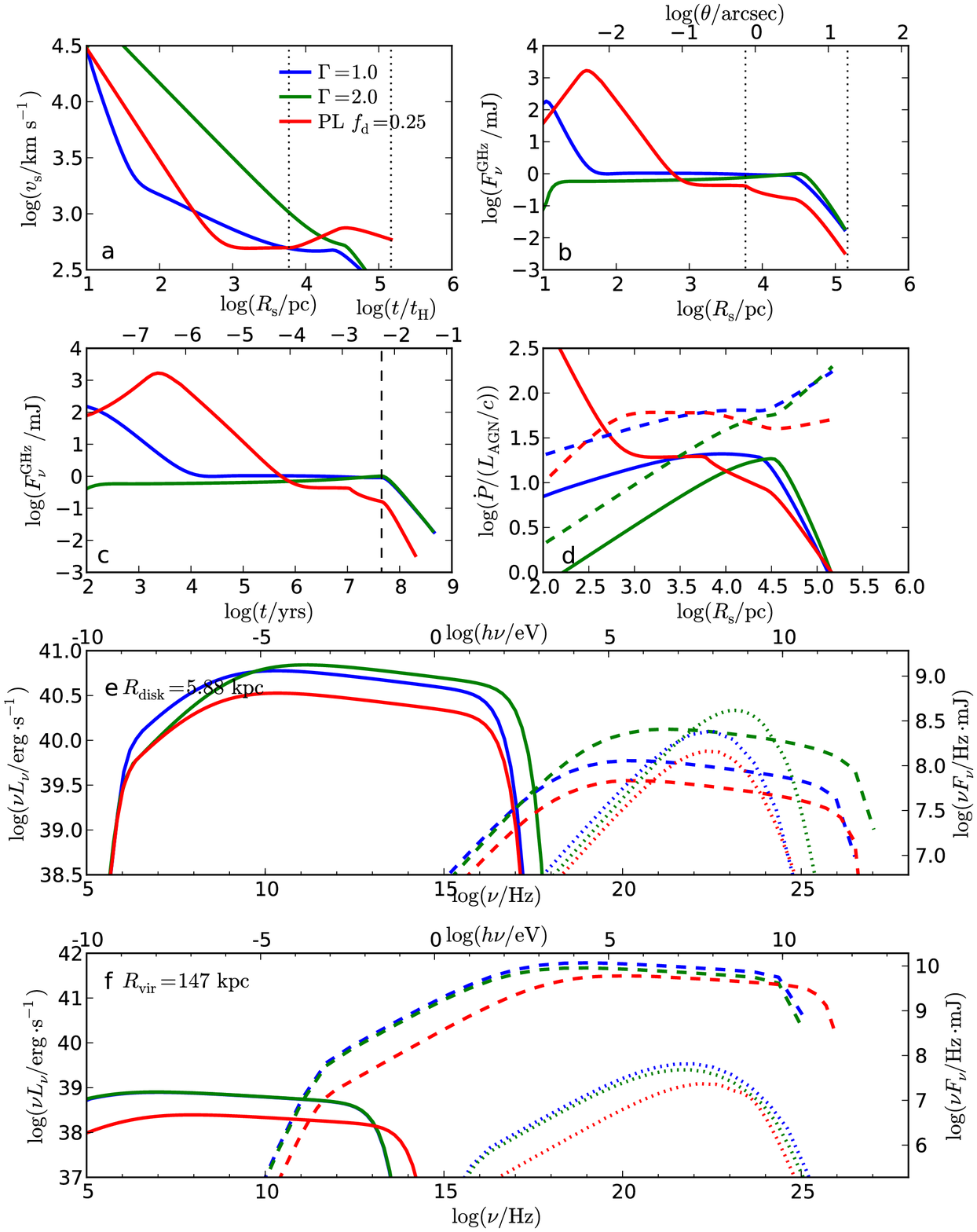}
\vglue -0.5cm
\caption{\label{gc}Dependence of outflow hydrodynamics and emission on gas density profile of galaxy clusters.
We fix $\mh=10^{12}\msun$ and $z=1.0$. We compare galaxy cluster gas density profile with the broken power-law profile ($\fd=0.25$).
The configuration and physical significance of the subplots are the same as Fig.~\ref{fd}.
}
\end{figure*}
\subsection{Energy or momentum conserving outflow}
\begin{figure}
\includegraphics[angle=0,width=\columnwidth]{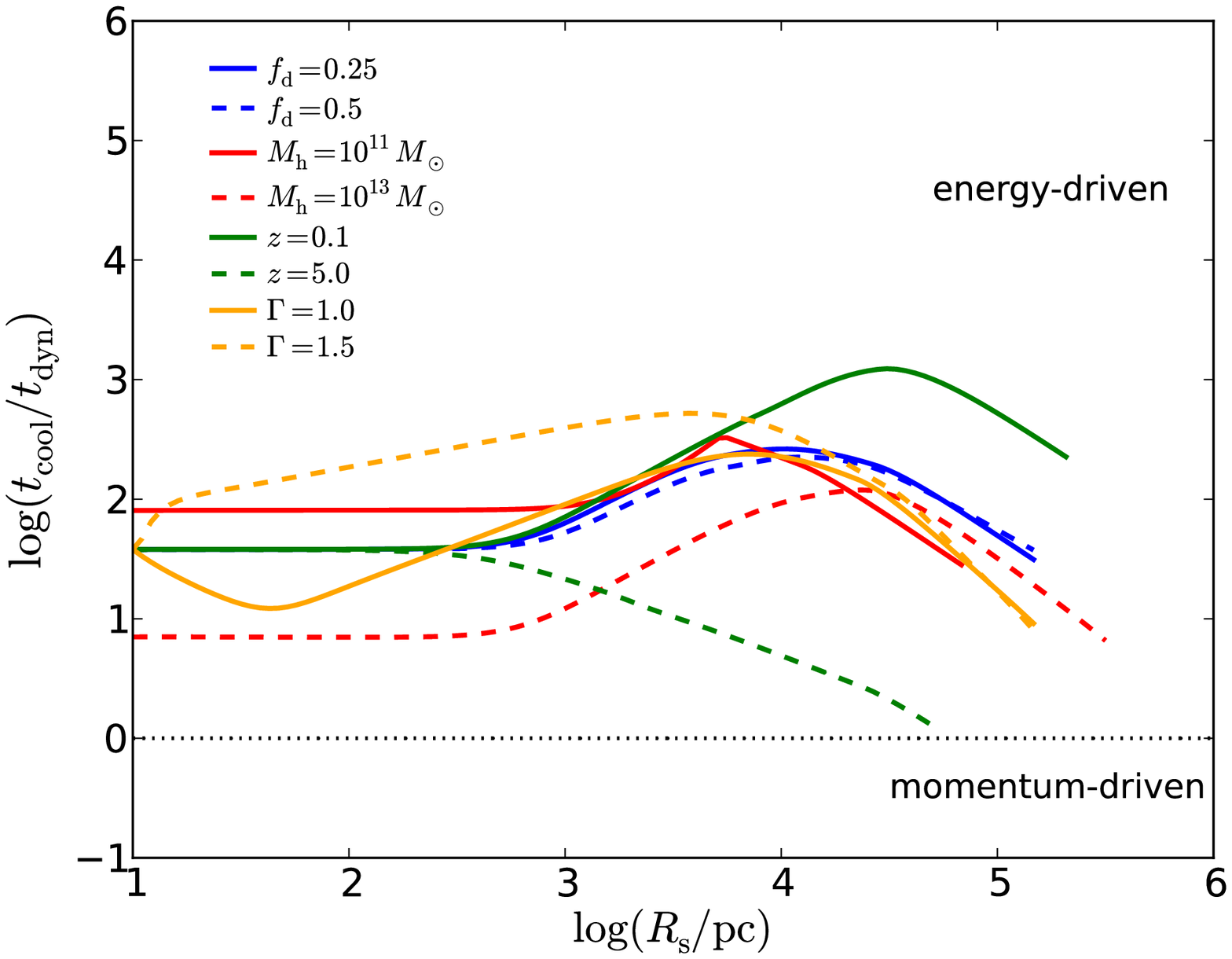}
\caption{\label{timescale}Ratio of radiative cooling timescale in the shocked wind region to outflow's dynamical time. The default values of the parameters are: $\fd=0.25$, $\mh=10^{12}\msun$ and $z=1.0$. Each line represents a specific parameter modified from its default value while the other parameters are fixed at the default values.
The dotted line separates the momentum and energy conserving regimes.
}
\end{figure}

Another important dynamics issue is whether the outflow is momentum or energy conserving.
In the momentum-driven regime, thermal energy in the shocked wind region is efficiently radiated away, while in energy-driven outflows, such radiative losses are insignificant.
We compare the timescale of the most efficient radiative cooling processes discussed in \S3 in the shocked wind, $\tcool$, with the dynamical timescale of the outflow, given by $\tdyn=\rs/\vs$, as shown in Fig. \ref{timescale}.

The plot shows $\tcool/\tdyn$ for several representative cases and indicates that for some cases
the outflow starts propagating as partially momentum-driven. Once the shell reaches $\sim 100\,\rm pc$, the partially momentum-driven regime breaks down and the shocked wind region no longer cools rapidly.
At larger radii, the soft photon energy density is dominated by CMB photons and $\tcool/\tdyn$ decreases consequently.
However, the energy conserving nature remains unchanged at larger radii, which is in agreement with recent observations \citep{2015Natur.519..436T} and theoretical calculations \citep{2012MNRAS.425..605F, 2012ApJ...745L..34Z}.

These results suggest that most of the wind kinetic energy
is converted to the kinetic energy of the outflow, giving $\dot{P}^2/\dot{\ms}\sim\dot{P}_{\rm rad}^2/\dot{M_{\rm in}}$,
where $\dot{P}_{\rm rad}=\lagn/c$ is the momentum flux of AGN's radiation field and $\dot{M_{\rm in}}$ is the mass injection rate of the wind from the central source \citep{2012ApJ...745L..34Z}.
We can write the momentum flux of the outflow normalized to AGN's radiation as $ \dot{P}/\dot{P}_{\rm rad}\sim\vin/\vs$. 
This relation is illustrated in panel \textit{d} of Figures 3-6.

%
\section{Conclusions and Discussion}
\label{sec:summary}
We study the hydrodynamics of AGN-driven outflows out to galactic halo scales and the resulting non-thermal emission from the fast forward outflow shock propagating into the ambient medium. 
We have found that the outflow decelerates rapidly to a nearly constant velocity of $\sim 10^3\,\kms$ within the galaxy disk and accelerates once it enters the halo until the central BH shuts off.
Around this time, the outflow can reach the edge of the halo.
We have verified that the outflow is energy-conserving on large radii, consistently with recent observations \citep{2015Natur.519..436T} and theoretical predictions \citep{2012MNRAS.425..605F, 2012ApJ...745L..34Z}.
The predicted non-thermal emission from outflow shocks in MW mass halos up to a redshift $z$ of $5$ is detectable over a broad range of wavelengths.
At $z\sim0.1$, the $2'$ angular scale emission is detectable by \textit{JVLA} and \textit{SKA} in radio band, \textit{ALMA} in mm/sub-mm band, \textit{JWST} and \textit{HST} in optical and infrared bands, marginally detectable in X-ray band by \textit{Chandra}, \textit{XMM-Newton} and \textit{ATHENA}.
At $z\sim1$, the signal remains observable in radio band and marginally detectable in infrared and optical bands with an angular scale of $\sim 18''$.
The detection is promising even at high redshifts ($z\sim5$) in the radio, infrared and optical bands with an angular scale of $\sim 8''$.
For lower mass halos the detection should limit within the local Universe.

We find that the detailed gas distributions do not significantly affect the hydrodynamics of the outflow while the halo mass plays a more important role in regulating the outflow dynamics.
We show a near universality of the non-thermal emission within the galaxy disk for different gas distributions of galaxies with same halo masses, which breaks down on the halo scale as a result of distinct density profile for tenuous halo gas.
The halo mass determines the intensity of the emission since the BH mass is scaled self-consistently with halo mass.
Consequently, non-thermal emission from outflows embedded in low mass halos is $\sim 1-3$ orders of magnitude fainter than that in MW mass halos.
We conclude that the halo mass is the dominant factor in regulating the dynamics and emission of the outflow.
In order to distinguish between different gas density distributions, halo scale observations are required.

The predicted non-thermal emission should be an observational signature of the existence of extended gas in galaxy halos in a wide range of redshifts.
With this method, one can probe the evolution of gaseous halos at early cosmic times.
Thermal X-ray emission from free-free cooling at the forward wind shock was proposed to be an observational signature of kpc-scale outflows powered by AGN \citep{2015MNRAS.447.3612N}. 
The predicted thermal X-ray luminosity at 1keV band is smaller than our non-thermal X-ray prediction and dinimishes with increasing outflow shock radius given our assumption about the gas density profile in the galaxy and halo.
Since the travel time of the outflow shocks is comparable to AGN's lifetime, most of the detected halos still host an active quasar, targets can be selected for observations as an AGN.
On the other hand, subtraction of the much brighter emission from the AGN is required to measure the extended diffuse emission from the outflow shocks.
Radio interferometry can resolve the luminous central source and subtract emission from it to obtain the extended emission on halo scale.
For optical and infrared observations, the extended emission can be subtracted using techniques similar to the removal of quasar light in \textit{HST} images \citep{1995ApJ...454L..77M, 1997ApJ...479..642B}.

A source of contamination to the extended non-thermal emission is the scattered quasar light by the surrounding electrons in the halo \citep{1990ApJ...363..344W, 2000MNRAS.312..567Y, 2004ApJ...602..659H}.
We find that the optical depth for Thomson scattering through the halo is $\sim 10^{-5}$, so $\sim 10^{-5}$ of the observed flux from the AGN is expected to diffuse throughout the halo.
For a $10^{12}\,\msun$ mass halo, the bolometric luminosity of the scattered radiation is $\sim 10^{40}\,\ergs$, which is comparable to the non-thermal emission at infrared and optical frequencies from outflow shocks in halos within $z\lesssim 1$ and negligible for halos at $z\sim5$.
One possible way to distinguish the scattered radiation from the non-thermal emission is by polarimetric measurement.
Additionally, the scattered light is diffused throughout the halo at any given time while the emission from outflow shocks shows a discontinuity at the shock front.
As the outflow propagates farther into the halo, the scattered quasar light no longer exists as the quasar fades away.
There is no contamination from scattered AGN photons in radio band from radio-quiet quasars, which takes up $\sim 90\%$ of the population, so the non-thermal emission can be more easily identified in radio wavelength \citep{2015MNRAS.447.3612N}.
Therefore, radio observation is expected to be most effective in detecting the halo scale non-thermal emission from outflows in a wide range of redshifts.
It should be noted that the predicted radio emission from outflow shocks exists without the presence of relativistic jets, which account for the radio emission from radio galaxies \citep{2014ARA&A..52..589H}.

There are a few uncertainties in our model. First, spherical symmetry of both gas distribution and outflow shell is likely to be unrealistic. 
In fact, the outflow may be collimated from the start or can propagate along the path of least resistance, forming a bipolar or bicone structure. 
Observations of kpc-scale molecular outflows suggest a wide-angle biconical geometry \citep{2011ApJ...729L..27R, 2015arXiv150301481F}.
Biconical outflows with small opening angle could have less impact on the ambient medium.
Second, the detailed gas distribution is uncertain and can be complicated by galaxy-to-galaxy variations, which can greatly dependend on galaxy types as well as the specific gas phase.
Finally,  we find that the terminal velocity of the outflow arriving at the edge of the halo is $\sim 10^3\,\kms$, which is still large enough for farther propagation of the outflow into the IGM.
The propagation dynamics of the outflow into the IGM is beyond the scope of this paper.
Along some directions gas accretion onto the galaxy could impede the developing outflow \citep{2015MNRAS.448..895S}.
%
\section*{Acknowledgements}
We thank Mark Reid and Lorenzo Sironi for helpful comments on the manuscript.
This work was supported in part by NSF grant AST-1312034.

%
\label{lastpage}
\end{document}